\documentclass[12pt]{article}
\usepackage{amssymb,textcomp,amsmath,graphicx}
\usepackage{epstopdf}
\long\def\symbolfootnote[#1]#2{\begingroup
\def\thefootnote{\fnsymbol{footnote}}
\footnote[#1]{#2}\endgroup}
\def\lsim{\:\raisebox{-0.75ex}{$\stackrel{\textstyle<}{\sim}$}\:}
\def\gsim{\:\raisebox{-0.75ex}{$\stackrel{\textstyle>}{\sim}$}\:}

\numberwithin{equation}{section}

\begin{document}

\begin{center}
 \Large{Running-Mass Inflation Model and Primordial Black Holes}
\end{center}
\begin{center}
\large{Manuel Drees$^{1,~2~*}$} and \large{Encieh Erfani$^{1,~\dag}$}
\end{center}
\begin{center}
 \textit{$^1$Physikalisches Institut and Bethe Center for Theoretical Physics, 
 Universit\"{a}t Bonn,\\ Nussallee 12, 53115 Bonn, Germany \\ and \\
$^2$School of Physics, KIAS, Seoul 130--722, Korea}
\end{center}

\date{}

\symbolfootnote[0]{$^{*}$drees@th.physik.uni-bonn.de}
\symbolfootnote[0]{$^{\dag}$erfani@th.physik.uni-bonn.de}

\begin{abstract}
We revisit the question whether the running--mass inflation model allows the
formation of Primordial Black Holes (PBHs) that are sufficiently long--lived
to serve as candidates for Dark Matter. We incorporate recent cosmological
data, including the WMAP 7--year results. Moreover, we include ``the running
of the running'' of the spectral index of the power spectrum, as well as the
renormalization group ``running of the running'' of the inflaton mass
term. Our analysis indicates that formation of sufficiently heavy, and hence
long--lived, PBHs still remains possible in this scenario. As a by--product,
we show that the additional term in the inflaton potential still does not
allow significant negative running of the spectral index.
\end{abstract}
\newpage


\section{Introduction}

\paragraph{}

The Cosmic Microwave Background (CMB) is very smooth. Full--sky observations
allow to expand the measured CMB temperature in spherical harmonics $Y_{\ell
  m}$. One can then determine the size of the anisotropies as a function of
$\ell$, with larger $\ell$ corresponding to smaller angles, and hence smaller
length scales. Current CMB observation probe this power spectrum down to
(comoving) length scales of about one Mpc. These observations imply very small
primordial density perturbations at such large length scales, characterized by
the power $\mathcal{P}_{\mathcal{R}_{c}} \simeq 10^{-9}$.

However, it is possible that the primordial density perturbations become much
larger at smaller length scales, beyond the range probed by cosmological
observations. Indeed, it is conceivable that these perturbations are so large
that overdense regions collapse to form Primordial Black Holes (PBHs) just
after the end of inflation \cite{Hawking1, Carr1, Carr2}. They are called
``primordial'' since they do not originate from the gravitational collapse of
burnt--out stars; they could thus have any mass, including masses well below
or well above stellar masses. Here we assume that the fluctuations which
sourced the PBHs are also generated during inflation, specifically towards the
end of inflation, well after the length scales probed by conventional
cosmological observations exited the horizon.

There are various constraints on PBH formation. For example, the density of
roughly stellar mass black holes has to satisfy limits from searches for
microlensing. Very light black holes could have evaporated (via Hawking
radiation) in the epoch of Big Bang nucleosynthesis, altering the predicted
isotope abundances. These and other constraints have recently been compiled in
\cite{Green}. They can be translated into upper limits on the amplitude of the
power spectrum at the length scales relevant for PBH formation, typically
$\mathcal{P}_{\mathcal{R}_{c}} < 10^{-2}-10^{-1}$ with some scale dependence.

Clearly the power spectrum has to change dramatically towards the end of
inflation for PBH formation to occur. In the framework of standard slow--roll
inflation, this implies that the slow--roll parameters, and hence the inflaton
potential, also should show large variations. One simple, and yet well
motivated, model that can feature large variations of the slow--roll
parameters is the running-mass model \cite{Stewart1, Stewart2}, a type of
inflationary model which emerges naturally in the context of supersymmetric
extensions of the Standard Model. The model is of the single--field type, but
nevertheless it has relatively strong scale--dependence (running) of the
spectral index. Previous studies of this model \cite{Leach} showed that it can
indeed accommodate sufficiently large positive running of the spectral index
to allow for PBH formation. However, the recent 7--year WMAP data
\cite{WMAP7} prefer {\em negative} running of the spectral index. 

In this paper we therefore revisit the running mass model. We expand on
refs.\cite{Leach} by including the running of the running of the spectral
index, which is only weakly constrained by existing CMB data. Similarly, we
include the running of the running of the inflaton mass parameter in the
potential. We show that for some (small) range of parameters, this model can
still accommodate sufficiently large density perturbations near the end of
inflation to allow for the formation of PBHs with mass larger than $10^{15}$
g, which are sufficiently long--lived to be candidates for Cold Dark Matter
(CDM). As a by--product we show that even with the additional term in the
potential, the model does not allow for significant {\em negative} running of
the spectral index.

The remainder of this paper is organized as follows: In section 2 we briefly
review the formalism of PBH formation. In section 3 we discuss the
running--mass inflation model, including the running of the running of the
mass. We compute the slow--roll parameters which allow us to determine the
spectral index, its running, and the running of the running. In section 4 we
perform a numerical analysis where we compute the spectral index at PBH scales
exactly, by numerically solving the equation of motion of the inflaton field.
Finally, we conclude in section 5.

\section{Formation of Primordial Black Holes}
\paragraph{}

PBHs may have formed during the very early universe, and if so can have
observational implications at the present epoch, e.g. from effects of their
Hawking evaporation \cite{Hawking1} for masses $\lsim 10^{15}$ g, or by
contributing to the present ``cold'' dark matter density if they are more
massive than $10^{15}$ g \cite{Carr2}. (These PBHs would certainly be massive
enough to be dynamically ``cold''\footnote{It has been argued
  \cite{remnant} that BH evaporation might leave a stable remnant with mass of
  order of the Planck mass, which could form CDM; we will not pursue this
  possibility here.}).

The traditional treatment of PBH formation is based on the Press-Schechter
formalism \cite{Press-Schechter} used widely in large--scale structure
studies. Here the density field is smoothed on a scale $R(M)$. In the case at
hand, $R(M)$ is given by the mass enclosed inside radius $R$ when $R$ crossed
the horizon. The probability of PBH formation is then estimated by simply
integrating the probability distribution $P(\delta;R)$ over the range of
perturbations $\delta$ which allow PBH formation: $\delta_{\rm th} < \delta
< \delta_{\rm cut}$, where the upper limit arises since very large
perturbations would correspond to separate closed 'baby' universes
\cite{Carr2, Hawking2}. We will show that in practice $P(\delta;R)$ is such a
rapidly decreasing function of $\delta$ above $\delta_{\rm th}$ that the upper
cutoff is not important. The threshold density is taken as $\delta_{\rm th} >
w$, where $w = p/\rho$ is the equation of state parameter describing the epoch
during which PBH formation is supposed to have occurred \cite{Carr2}. Here we
take $w=1/3$, characteristic for the radiation dominated epoch which should
have started soon after the end of inflaton. However the correct value of the
threshold $\delta_{\rm th}$ is quite uncertain. Niemeyer and Jedamzik
\cite{Jedamzik} carried out numerical simulations of the collapse of the
isolated regions and found the threshold for PBH formation to be $0.7$. We
will show that PBHs abundance is sensitive to the value of $\delta_{\rm th}$.

The fraction of the energy density of the Universe in which the density
fluctuation exceeds the threshold for PBH formation when smoothed on scale
$R(M)$, $\delta(M) > \delta_{\rm th}$, which will hence end up in PBHs with
mass $\geq \gamma M$,\footnote{Throughout we assume for simplicity that the PBH mass is a fixed fraction $\gamma$ of the horizon mass corresponding to the smoothing scale. This is not strictly true. In general the mass of PBHs is expected to depend on the amplitude, size and shape of the perturbations \cite{Jedamzik,shape}.} is given as in Press--Schechter theory by\footnote{We follow ref.\cite{factor2} in including a factor of two on the right--hand side; this factor is not very important for delineating the inflationary model parameters allowing significant PBH formation. Moreover, we set $\delta_{\rm cut}$ to infinity.}
\begin{equation} \label{fofm}
f( \geq M) = 2 \gamma \int _{\delta_{\rm th}}^{\infty} P(\delta; M(R))
\text{d}\delta\,. 
\end{equation} 
Here $P(\delta; M(R))$ is the probability distribution function (PDF) of the
linear density field $\delta$ smoothed on a scale $R$, and $\gamma$ is the
fraction of the total energy within a sphere of radius $R$ that ends up inside
the PBH. 

For Gaussian fluctuations, the probability distribution of the smoothed
density field is given by\footnote{This PDF is often written as
  $P(\delta(R))$. However, we think it is more transparent to consider $P$ to
  be the PDF of $\delta$, which is just an integration variable in
  eq.(\ref{fofm}). Eq.(\ref{pofr}) shows that the functional form of
  $P(\delta)$ depends on the parameter $R$, which in turn depends on the
  horizon mass $M$.}
\begin{equation} \label{pofr}
P(\delta;R) = \dfrac {1} {\sqrt{2\pi} \sigma_{\delta}(R)} \exp\left( -\dfrac
{\delta^{2}} {2\sigma_{\delta}^{2}(R)} \right)\,.
\end{equation}
This PDF is thus uniquely determined by the variance $\sigma_{\delta}(R)$ of
$\delta$, which is given by
\begin{equation} \label{sigma}
\sigma_{\delta}^{2}(R) = \int_{0}^{\infty} W^{2}(kR) \mathcal{P}_{\delta}(k)
\dfrac{\text{d}k} {k}\,.
\end{equation}
In order to compute the variance, we therefore have to know the power spectrum
of $\delta$, $\mathcal{P}_{\delta}(k) \equiv k^3/ (2 \pi^2) \left \langle
|\delta_k|^2 \right \rangle$, as well as the volume--normalized Fourier
transform of the window function used to smooth $\delta$, $W(kR)$. 

It is not obvious what the correct smoothing function $W(kR)$ is; a top--hat
function has often been used in the past, but we prefer to use a Gaussian
window function\footnote {Bringmann \textit{et al.} \cite{Bringmann} argued
  that a top--hat window function predicts a larger PBH abundance.},
\begin{equation} \label{w}
 W(kR) = \exp\left(-\dfrac{k^{2}R^{2}}{2} \right)\,. 
\end{equation}

Finally, on comoving hypersurfaces there is a simple relation between the
density perturbation $\delta$ and the curvature perturbation $\mathcal{R}_{c}$
\cite{Lyth Book}: 
\begin{equation} \label{delta}
\delta(k,t) = \dfrac {2(1+w)} {5+3w} \left( \dfrac{k}{aH} \right)^2
\mathcal{R}_{c}(k)\,,
\end{equation}
The density and curvature perturbation power spectra are therefore related by
\begin{equation} \label{pofk}
\mathcal{P}_{\delta}(k,t) = \dfrac{4(1+w)^2} {(5+3w)^2} \left( \dfrac{k}{aH}
\right)^4 \mathcal{P}_{\mathcal{R}_{c}}(k)\,.
\end{equation}

The mass fraction of the Universe that will collapse into PBHs can now be
computed by inserting eqs.(\ref{pofk}) and (\ref{w}) into eq.(\ref{sigma}) to
determine the variance as function of $R$. This has to be used in
eq.(\ref{pofr}), which finally has to be inserted into eq.(\ref{fofm}). Since
we assume a Gaussian $P(\delta)$ in eq.(\ref{pofr}), the integral in eq.(\ref{fofm}) simply gives an error function.

In order to complete this calculation one needs to relate the mass $M$ to the
comoving smoothing scale $R$. The number density of PBHs formed during the
reheating phase just after the end of inflation will be greatly diluted by the
reheating itself. We therefore only consider PBHs which form during the
radiation dominated era after reheating is (more or less) complete. The
initial PBHs mass $M_{\text{PBH}}$ is related to the particle horizon mass
$M$ by\footnote{Throughout the paper we put $c=\hbar=k_{B}=1.$}
\begin{equation}
 M_{\text{PBH}}=\gamma M = \dfrac{4\pi}{3}\gamma \rho H^{-3}\, ,
\end{equation}
when the scale enters the horizon, $R=(aH)^{-1}$. Here the coefficient
$\gamma$, which already appeared in eq.(\ref{fofm}), depends on the details of
gravitational collapse. A simple analytical calculation suggest that $\gamma
\simeq w^{3/2} \simeq 0.2$ during the radiation era \cite{Carr2}. During
radiation domination $aH\varpropto a^{-1}$, and expansion at constant entropy
gives $\rho \varpropto g_{*}^{-1/3}a^{-4}$ \cite{Turner Book} (where $g_{*}$
is the number of relativistic degrees of freedom, and we have approximated the
temperature and entropy degrees of freedom as equal). This implies that
\begin{equation}
M_{\text{PBH}} = \gamma M_{\text{eq}} (k_{\text{eq}}R)^2 \left( \dfrac
{g_{\ast,\text{eq}}} {g_{\ast}} \right)^{1/3}\, , 
\end{equation}
where the subscript ``eq'' refers to quantities evaluated at matter--radiation
equality. In the early Universe, the effective relativistic degree of freedom
$g_{\ast}$ is expected to be of order $100$, while $g_{\ast,\text{eq}} = 3.36$
and $k_{eq} = 0.07 \Omega_{\text{m}} h^{2}\ \text{Mpc}^{-1} (\Omega_{\text{m}}
h^{2} = 0.1334$ \cite{WMAP7}). The horizon mass at matter-radiation equality
is given by
\begin{equation}
M_{\text{eq}} = \dfrac {4\pi} {3} \rho_{\text{rad,eq}} H_{\text{eq}}^{-3} =
\dfrac {4\pi} {3} \dfrac {\rho_{\text{rad},0}} {k_{\text{eq}}^3 a_{\text{eq}}}\,,
\end{equation}
where $a_{\text{eq}}^{-1}=(1+z_{\text{eq}})=3146$ and (assuming three species
of massless neutrinos) $\Omega_{\text{rad},0}h^{2}=4.17\times 10^{-5}$. Then
it is straightforward to show that 
\begin{equation} \label{R}
\dfrac {R} {1\ \text{Mpc}} = 5.54 \times 10^{-24} \gamma^{-\frac{1}{2}} \left(
\dfrac {M_{\text{PBH}}} {1\ \text{g}} \right)^{1/2} \left( \dfrac {g_{\ast}}
       {3.36} \right)^{1/6}\, . 
\end{equation}
Note that $M_{\rm PBH} \propto R^2$, not $\propto R^3$ as one might naively
have expected. Recall that $M_{\rm PBH}$ is related to the horizon mass at the
time when the comoving scale $R$ again crossed into the horizon. Larger scales
re--enter later, when the energy density was lower; this weakens the
dependence of $M_{\rm PBH}$ on $R$. Moreover, the lightest black holes to form
are those corresponding to a comoving scale that re--enters the horizon
immediately after inflation.\footnote{In fact, PBH formation might also occur
  on scales that never leave the horizon \cite{earlyPBH}. We do not consider
  this contribution here.}

The Gaussian window function in eq.(\ref{sigma}) strongly suppresses
contributions with $k > 1/R$. At the same time, the factor $k^4$ in
eq.(\ref{pofk}) suppresses contributions to the integral in eq.(\ref{sigma})
from small $k$. As a result, this integral is dominated by a limited range of
$k-$values near $1/R$. Over this limited range one can to good approximation
assume a power--law primordial power spectrum with {\em fixed} power
$n_S$\footnote{``Fixed'' here means that $n_S$ does not depend on $k$; however,
  $n_S$ does depend on $R$, since a large range of values of $R$ has to be
  considered for PBH formation of different masses.}
$\mathcal{P}_{\mathcal{R}_{c}}(k) = \mathcal{P}_{\mathcal{R}_{c}} (k_R)
(k/k_R)^{n_S(R)-1}$, with $k_R = 1/R$. With this ansatz, the variance of
the primordial density field at horizon crossing is given by
\begin{equation} \label{sigmaR}
\sigma_{\delta}^2 (R) = \dfrac{2(1+w)^2} {(5+3w)^2}
\mathcal{P}_{\mathcal{R}_{c}}(k_R) \Gamma[(n_S(R)+3)/2] \,,
\end{equation}
for $n_S(R)>-3$.

The power $\mathcal{P}_{\mathcal{R}_c}$ is known accurately at CMB scales; for
example, $\mathcal{P}_{\mathcal{R}_c}(k_0) = (2.43 \pm 0.11) \times 10^{-9}$
at the ``COBE scale'' $k_0 = 0.002$ Mpc$^{-1}$ \cite{WMAP7}. In order to
relate this to the scales relevant for PBH formation, we parameterize the
power spectrum as 
\begin{equation} \label{eqn}
\mathcal{P}_{\mathcal{R}_c}(k_R) = \mathcal{P}_{\mathcal{R}_c}(k_0)
(k_R/k_0)^{n(R)-1}\, . 
\end{equation}
It is important to distinguish between $n_S(R)$ and $n(R)$ at this
point. $n_S(R)$ describes the {\em slope} of the power spectrum at scales $k
\sim k_R = 1/R$, whereas $n(R)$ fixes the {\em normalization} of the spectrum
at $k_R \gg k_0$. The two powers are identical if the spectral index is
strictly constant, i.e. if neither $n_S$ nor $n$ depend on $R$. However, in this
case CMB data imply \cite{WMAP7} that $n = n_S$ is close to
unity. Eqs.(\ref{sigmaR}) and (\ref{eqn}) then give a very small variance,
leading to essentially no PBH formation.

Significant PBH formation can therefore only occur in scenarios with running
spectral index. We parameterize the scale dependence of $n$ as \cite{Kosowsky}:
\begin{equation} \label{nofr}
 n(R) \quad = \quad n_S (k_0) - \frac{1}{2!} \, \alpha_S\, \ln \left( k_0
 R \right) + \dfrac {1} {3!} \, \beta_S\, \ln^2 \left( k_0 R \right) +\dots\, ;
\end{equation}
recall that we are interested in $R \ll 1/k_0$, i.e. $\ln(k_0 R) < 0$.
The parameters $\alpha_S$ and $\beta_S$ denote the running of the effective
spectral index $n_S$ and the running of the running, respectively:
\begin{eqnarray} \label{runpar}
n_S (k_0) \quad &\equiv& \quad \left. \dfrac {d\ln
  \mathcal{P}_{\mathcal{R}_c}} {d \ln k} \right|_{k=k_0}\, , 
\nonumber \\ 
\alpha_S (k_0) \quad & \equiv &\quad \left. \dfrac {d n_S} {d\ln k}
\right|_{k=k_0} \, , 
\nonumber \\
\beta_S (k_0) \quad & \equiv & \quad \left. \dfrac {d^2 n_S} {d\ln^2
  k}\right|_{k=k_0}\,.  
\end{eqnarray}

Eq.(\ref{nofr}) illustrates the difference between $n(R)$ and $n_S(R)$. The
latter has an expansion similar to eq.(\ref{nofr}), but with the usual
Taylor--expansion coefficients, $1$ in front of $\alpha_S$ and $1/2$ in front
of $\beta_S$. One therefore has
\begin{equation} \label{nsofn}
n_S(R)\quad = \quad n(R) - \frac {1}{2} \, \alpha_S\, \ln \left( k_0 R \right) +
\frac{1}{3} \, \beta_S \, \ln^2 \left( k_0 R \right) + \dots \, .
\end{equation}
Setting $n_S(k_0)=1$ for simplicity, eq.(\ref{nsofn}) implies $n_S(R) = 2
n(R) - 1$ for $\beta_S = 0$, and $n_S(R) = 3 n(R) - 2$ for $\alpha_S = 0$. We
will compute the variance $\sigma(R)$, and hence the PBH fraction $f$, for
these two relations; they represent extreme cases if neither $\alpha_S$ nor
$\beta_S$ is negative.

\begin{figure}[h!]
\centering
\includegraphics[angle=270,width=1.\textwidth]{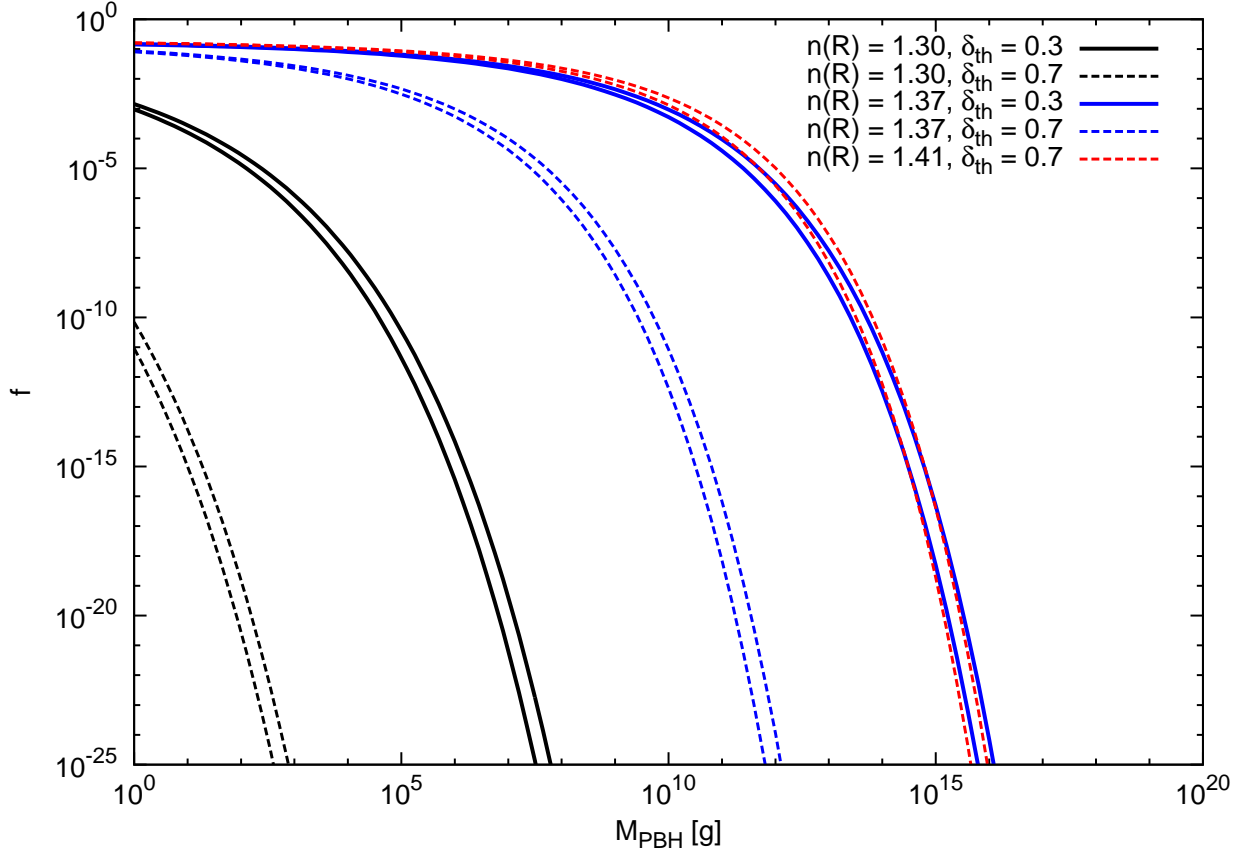}
\caption{Fraction of the energy density of the universe collapsing into PBHs
 as a function of the PBH mass, for three different values of $n(R)$ and two
 different choices of the threshold $\delta_{\rm th} = 0.3 \ (0.7)$ for the
 solid (dashed) curves. On the upper [lower] of two curves with equal pattern
 and color we have assumed $n_S(R) = 2 n(R)-1$ [$n_S(R) = 3 n(R) - 2$].}
\label{fig:Fig1}
\end{figure}

The result of this calculation is shown in figure~\ref{fig:Fig1}. Here we have
fixed $\gamma = 0.2$, and show results for two choices of the threshold
$\delta_{\rm th}$ and three choices of $n(R)$. We see that scenarios where
$n(R) = 1.3$ (or smaller) are safe in the SM, because there is no
model--independent limit on $f$ for $M_{\rm PBH} < 10^{10}$ g \cite{Green}. As
noted earlier, PBHs contributing to Dark Matter today must have $M_{\rm PBH}
\gsim 10^{15}$ g; at this mass, they saturate the DM relic density if $f
\simeq 5\times10^{-19}$.\footnote{Note that $f$ describes the fraction of the
  energy density in PBHs at the time of their formation. Since they behave
  like matter at all times, their fractional contribution to the energy
  density increases during the radiation--dominated epoch, and stays
  essentially constant during the subsequent matter--dominated epoch.}
Figure~\ref{fig:Fig1} shows that this requires $n(R) \simeq 1.37 \ (1.41)$ for
$\delta_{\rm th} = 0.3 \ (0.7)$. The dependence on $n_S$ is much milder.

Figure~\ref{fig:Fig1} also illustrates a serious problem that all scenarios that
aim to explain the required CDM density in terms of post--inflationary PBH
formation face. We just saw that this can happen only if the spectral index
$n$ increases significantly between the scales probed by the CMB and other
cosmological observations and the scale $R \simeq 10^{-9}$ pc relevant for the
formation of $10^{15}$ g PBHs. However, $n$ must then {\em decrease} rapidly
when going to slightly smaller length scales, since otherwise one would {\em
  overproduce} lighter PBHs. For example, successful Big Bang Nucleosynthesis
requires \cite{Green} $f(10^{13} \ {\rm g}) \leq 2 \times 10^{-20}$, about 12
orders of magnitude below that predicted by keeping $n(R)$ fixed at the value
required for having $10^{15}$ g PBHs as CDM candidates. We will come back to
this point at the end of our paper.

Another problem is that the expansion of $n(R)$ in eq.(\ref{nofr}) will
generally only be accurate if $|\ln(k_0 R)|$ is not too large. This is the
case for cosmological observations, which probe scales $\gsim 1$ Mpc. The
expansion becomes questionable for the scales probed by PBH formation. For
example, fixing $k_0 = 0.002$ Mpc$^{-1}$, eq.(\ref{R}) gives $|\ln (k_0R)| =
41.1$ for $M_{\rm PBH} = 10^{15}$ g. Fortunately within the framework of a
given inflationary scenario this second problem can be solved by computing
$n(R)$ and $n_S(R)$ exactly, rather than using the expansion (\ref{nofr}).

Observational bounds on $n_S$ and $\alpha_S$ can e.g. be found in
ref.\cite{WMAP7}. By requiring that these parameters remain within the $2
\sigma$ ranges for all scales $k$ between $10^{-4}$ Mpc$^{-1}$ and $10$
Mpc$^{-1}$, we find $\beta_S \lsim 0.03$. Here we used the error bars derived
from the analysis of WMAP7 data plus data on baryonic acoustic oscillations
(BAO), ignoring possible tensor modes (as appropriate for small--field
models), and including independent measurements of the Hubble constant as a
prior (i.e. we used the ``WMAP7+BAO+$H_0$'' data set). Here we use the pivot
scale $k_0 = 0.0155$ Mpc$^{-1}$ where $n_S(k_0)$ and $\alpha_S(k_0)$ are
uncorrelated. On the other hand, eq.(\ref{nofr}) shows that for $\alpha_S = 0$
we only need $\beta_S(k_0) \simeq 0.0016$ in order to generate sufficiently
large density perturbations to allow formation of $10^{15}$ g PBHs. Even if we
set $\alpha_S(k_0)$ equal to its central value, $\alpha_S(k_0) = -0.022$, we
only need $\beta_S(k_0) \simeq 0.0032$. Including the running of the running
of the spectral index thus easily allows to accommodate PBH formation in
scenarios that reproduce all current cosmological observations at large
scales.

Of course, this kind of model--independent analysis does not show whether
simple, reasonably well--motivated inflationary models exist that can generate
a sufficiently large $\beta_S$. One scenario that quite naturally
accommodates strong scale dependence of the spectral index is the
running--mass inflation model \cite{Stewart1, Stewart2}. We now apply our
results to this model.

\section{Running-Mass Inflation Model}

\paragraph{}

This model, proposed by Stewart \cite{Stewart1, Stewart2}, exploits the
observation that in field theory the parameters of the Lagrangian are scale
dependent. This is true in particular for the mass of the scalar inflaton,
which can thus be considered to be a ``running'' parameter.\footnote{Note that the physical, or pole, mass of the inflaton is not ``running''; however, at the quantum level the physical mass differs from the parameter $m_\phi$ appearing in eq.(\ref{pot1}) even if $\phi = 0$.} The running of the mass
parameter can be exploited to solve the ``$\eta-$problem'' of inflation in
supergravity \cite{lr}. This problem arises because the vacuum energy driving
inflation also breaks supersymmetry. In ``generic'' supergravity models the
vacuum energy therefore gives a large (gravity--mediated) contribution to the
inflaton mass, yielding $|\eta| \sim 1$. However, this argument applies to the
scale where SUSY breaking is felt, which should be close to the (reduced)
Planck scale $M_{\rm P} = 2.4\times10^{18}$ GeV. In the running mass model,
renormalization group (RG) running of the inflaton mass reduces the inflaton
mass, and hence $|\eta|$, at scales where inflation actually happens. There
are four types of model, depending on the sign of the squared inflaton mass at
the Planck scale, and on whether or not that sign change between $M_{\rm P}$
and the scales characteristic for inflation \cite{Covi and Lyth}.

The simplest running--mass model is based on the inflationary potential
\begin{equation} \label{pot1}
V_\phi = V_0 + \dfrac {1} {2} m_\phi^2(\phi) \phi^2\,,
\end{equation}
where $\phi$ is a real scalar; in supersymmetry it could be the real or
imaginary part of the scalar component of a chiral superfield. The natural
size of $|m^2_\phi(M_P)|$ in supergravity is of order $V_0/M_{\rm P}^2$. Even
for this large value of $m_\phi^2$, which gives $|\eta| \sim {\cal O}(1)$, the
potential will be dominated by the constant term for $\phi^2 \ll M_{\rm P}^2$;
as mentioned above, running is supposed to reduce $|m^2_\phi|$ even more at
lower $\phi^2$.

The potential (\ref{pot1}) would lead to eternal inflation. One possibility to
end inflation is to implement the idea of hybrid inflation \cite{Hybrid
  Linde}. To that end, one introduces a real scalar ``waterfall field''
$\psi$, and adds to the potential the terms\footnote{The parameters of this
  potential will in general also be scale dependent; however, this is
  immaterial for our argument.}
\begin{equation} \label{pot2}
V_\psi = \frac {\lambda}{4} \phi^2 \psi^2 - \sqrt{ \frac {V_0 \kappa} {6} }
\psi^2 + \frac {\kappa} {24} \psi^4\,.
\end{equation}
Here $\lambda$ and $\kappa$ are real couplings, and the coefficient of the
$\psi^2$ term has been chosen such that $V_{\inf} = V_\phi + V_\psi$ has a
minimum with $\langle V \rangle = 0$ if $\phi=0, \ \psi \neq 0$. As long as
$\phi^2 > \sqrt{ \frac {8 V_0 \kappa}{3 \lambda^2} }$, $\psi$ remains frozen
at the origin. Once $\phi^2$ falls below this critical value, $\psi$ quickly
approaches its final vacuum expectation value, given by $\langle \psi
\rangle^2 = \sqrt{ \frac {24 V_0} {\kappa} }$, while $\phi$ quickly goes to
zero, thereby ``shutting off'' inflation. However, in this paper we focus on
the inflationary period itself; the evolution of perturbations at length
scales that left the horizon a few e--folds before the end of inflation should
not be affected by the details of how inflation is brought to an
end.\footnote{It has recently been pointed out that the waterfall phase might
  contribute to PBH formation \cite{new-Lyth}.}

During inflation, the potential is thus simply given by eq.(\ref{pot1}).
Here $m_{\phi}^2(\phi)$ is obtained by integrating an RG equation of the form
\begin{equation}
\dfrac {d m_\phi^2} {d\ln \phi} = \beta_m\,,
\end{equation}
where $\beta_m$ is the $\beta-$function of the inflaton mass parameter. If
$m_\phi^2$ is a pure SUSY--breaking term, to one loop $\beta_{m}$ can be
schematically written as \cite{Stewart1, Covi}
\begin{equation} \label{beta}
\beta_m =- \dfrac {2C} {\pi} \alpha \widetilde{m}^2 + \dfrac {D} {16\pi^{2}}
|\lambda_Y|^2 m_s^2\, ,
\end{equation}
where the first term arises from the gauge interaction with coupling $\alpha$
and the second term from the Yukawa interaction $\lambda_Y$.  $C$ and $D$
are positive numbers of order one, which depend on the representations of the
fields coupling to $\phi$, $\widetilde{m}$ is a gaugino mass parameter, while
$m_s^2$ is the scalar SUSY breaking mass--squared of the scalar particles
interacting with the inflaton via Yukawa interaction $\lambda_Y$.

For successful inflation the running of $m_\phi^2$ must be sufficiently strong
to generate a local extremum of the potential $V_\phi$ for some nonvanishing
field value, which we call $\phi_*$. The inflaton potential will obviously be
flat near $\phi_*$, so that inflation usually occurs at field values not very
far from $\phi_*$. We therefore expand $m_\phi^2(\phi)$ around $\phi =
\phi_*$. The potential we work with thus reads:
\begin{equation} \label{pot3}
V = V_0 + \dfrac {1} {2} m_\phi^2 (\phi_*) \phi^2 + \dfrac {1} {2} \ c\ \phi^2
\ \ln \left( \dfrac {\phi} {\phi_*} \right) + \dfrac {1} {4}
\ g\ \phi^2\ \ln^2 \left( \dfrac {\phi} {\phi_*} \right)\, .
\end{equation}
Here $c\equiv \left. \dfrac{d m_\phi^2}{d\ln \phi} \right|_{\phi=\phi_*}$ is
given by the $\beta-$function of eq.(\ref{beta}), and $g \equiv
\left. \dfrac{d^2 m_\phi^2} {d (\ln \phi)^2} \right|_{\phi=\phi_*}$ is given
by the scale dependence of the parameters appearing in eq.(\ref{beta}). In
contrast to earlier analyses of this model \cite{Covi and Lyth, Covi, clmo},
we include the $\ln^2(\phi/\phi_*)$ term in the potential. This is a two--loop
correction, but it can be computed by ``iterating'' the one--loop
correction.\footnote{There are also ``genuine'' two--loop corrections, which
  can not be obtained from a one--loop calculation, but they only affect the
  term linear in $\ln(\phi/\phi_*)$. They are thus formally included in our
  coefficient $c$.} Since the coefficient $g$ of this term is of fourth order
in couplings, one will naturally expect $|g| \ll |c|$. However, this need not
be true if $|c|$ ``happens'' to be suppressed by a cancellation in
eq.(\ref{beta}). Including the second correction to $m_\phi^2(\phi)$ seems
natural given that we also expanded the running of the spectral index to
second (quadratic) order.

Recall that we had defined $\phi_*$ to be a local extremum of $V_\phi$,
i.e. $V^\prime(\phi_*) = 0$. This implies \cite{Covi and Lyth} $m^2_* \equiv
m^2_\phi(\phi_*) = -\dfrac{1}{2}c$; this relation is not affected by the
two--loop correction $\propto g$.

In order to calculate the spectral parameters $n_S, \, \alpha_S$ and $\beta_S$
defined in eqs.(\ref{runpar}), we need the first four slow--roll parameters,
defined as \cite{Lyth Book}:\footnote{The powers on $\xi^2$ and $\sigma^3$ are
  purely by convention; in particular, $\xi^2$ could be negative.}
\begin{eqnarray} \label{slow-roll}
\epsilon & \equiv& \dfrac {M_{\rm P}^2} {2} \left( \dfrac {V^{\prime}} {V}
\right)^2 \, ,
\nonumber \\
\eta & \equiv & M_{\rm P}^2 \dfrac {V^{\prime\prime}} {V}\, , 
\nonumber \\
\xi^2 & \equiv & M_{\rm P}^4 \dfrac {V^{\prime} V^{\prime\prime\prime}} {V^2}\,,
\nonumber \\
\sigma^3 & \equiv& M_{\rm P}^6 \dfrac {V^{\prime2} V^{\prime\prime\prime\prime}}
      {V^3}\, .
\end{eqnarray}
where primes denote derivatives with respect to $\phi$. All these parameters
are in general scale--dependent, i.e. they have to be evaluated at the value
of $\phi$ that the inflaton field had when the scale $k$ crossed out of the
horizon. The spectral parameters are related to these slow--roll parameters by
\cite{Lyth Book}:
\begin{eqnarray} \label{nab}
n_S & = & 1 - 6 \epsilon + 2 \eta\, , 
\nonumber \\
\alpha_S & = & -24 \epsilon^2 + 16 \epsilon \eta - 2\xi^2 \, , 
\nonumber \\
\beta_S & = & -192 \epsilon^3 + 192 \epsilon^2 \eta - 32 \epsilon \eta^2 -24
\epsilon \xi^2 +2 \eta \xi^2 +2 \sigma^3 \, .
\end{eqnarray}

Slow--roll inflation requires $|\epsilon|, \, |\eta| \ll 1$. Combining
eqs.(\ref{pot3}) and (\ref{slow-roll}) we see that we need $V_0 \gg c \phi^2
L, \, g \phi^2 L^2$, where we have introduced the short--hand notation
\begin{equation} \label{l}
L \equiv \ln \frac {\phi} {\phi_*}\,.
\end{equation}
In other words, the inflaton potential has to be dominated by the constant
term, as noted earlier. Eqs.(\ref{slow-roll}) then imply two strong
inequalities between (combinations of) slow--roll parameters:
\begin{eqnarray} \label{hierarchies}
|\epsilon| &\ll& |\eta|\, ; 
\nonumber \\
|\epsilon \eta| & \ll & |\xi^2|\,.
\end{eqnarray}
The first relation means that $n_S-1$ is essentially determined by
$\eta$. Similarly, both relations together imply that $\alpha_S$ is basically
fixed by $\xi^2$, while only the last two terms in the expression for
$\beta_S$ are relevant; these two terms are generically of similar order of
magnitude. Finally, the factors of $V$ appearing in the denominators of
eqs.(\ref{slow-roll}) can be replaced by $V_0$. The spectral parameters are
thus given by:
\begin{eqnarray} \label{nab-slo}
n_S - 1 & = & 2 \frac{c M_{\rm P}^2} {V_0} \left[ L + 1 + \frac {g}{2c} \left(
  L^2 + 3 L + 1 \right) \right]\, ;
\nonumber \\
\alpha_S & = & -2 \left( \frac {c M_{\rm P}^2}{V_0} \right)^2 L \left[ 1 +
  \frac{g} {2c} \left( 2L + 3 \right) \right] \left[ 1 + \frac {g} {2c} \left(
  L + 1 \right) \right] \, ;\\
\beta_S & = & 2 \left( \frac {c M_{\rm P}^2}{V_0} \right)^3 L \left[ 1 + \frac
  {g}{2c} \left( L+1 \right) \right] \left[ 1 + \frac {g}{2c} \left( 3L+2
  \right) + \frac {g^2} {2c^2} \left( 3 L^2 + 5L + \frac{3}{2} \right) \right]
\nonumber \, , 
\end{eqnarray}
where $L$ has been defined in eq.(\ref{l}). Clearly the spectral index is not
scale--invariant unless $c$ and $g$ are very close to zero. Note that $V_0$
appears in eqs.(\ref{nab-slo}) only in the dimensionless combination $c M_{\rm
  P}^2 / V_0$, while $\phi$ only appears via $L$, i.e. only the ratio
$\phi/\phi_*$ appears in these equations.

In contrast, the absolute normalization of the power spectrum is given by (in
slow--roll approximation)
\begin{equation} \label{power}
\mathcal{P}_{\mathcal{R}_{c}} = \dfrac {1} {12\pi^2 M_{\rm P}^6} \dfrac
        {V^3}{V^{\prime2}} \, .
\end{equation}
This normalization is usually quoted at the ``COBE scale'' $k_0 = 0.002$
Mpc$^{-1}$, where $\mathcal{P}_{\mathcal{R}} = 2.43\times 10^{-9}$
\cite{WMAP7}. Applying our potential (\ref{pot3}) to eq.(\ref{power}),
replacing $V$ by $V_0$ in the numerator, we see that ${\cal P}_{{\cal R}_c}$
not only depends on $c M^2_{\rm P} / V_0$ and $L$, but also on the ratio
$V_0 / (M^2_{\rm P} \phi^2)$. We can thus always find parameters that give
the correct normalization of the power spectrum, for all possible combinations
of the spectral parameters.

We want to find out whether the potential (\ref{pot3}) can accommodate
sufficient running of $n_S$ to allow PBH formation. There are strong
observational constraints on $n_S$ and $\alpha_S$. It is therefore preferable
to use these physical quantities directly as inputs, rather than the model
parameters $c M^2_{\rm P} / V_0, \ L$ and $g/c$. To this end we rewrite the
first eq.(\ref{nab-slo}) as:
\begin{equation} \label {sol1}
\frac {c M^2_{\rm P}} {V_0} = \frac {n_S - 1} { 2 (L + 1) + \frac {g}{c}
  \left( L^2 + 3L + 1 \right) } \, .
\end{equation}
Inserting this into the second eq.(\ref{nab-slo}) gives:
\begin{equation} \label{sol2}
\alpha_S = - \frac { \left( n_S - 1 \right)^2 } {2} L \frac {\left[ 1 + \frac
    {g}{2c} \left( 2 L + 3 \right) \right]  \left[ 1 + \frac {g}{2c} \left(
    L+1 \right) \right] }
{ \left[ L + 1 + \frac {g} {2c} \left( L^2 + 3L + 1 \right) \right]^2 }\, .
\end{equation}
We thus see that the running of the spectral index is ``generically'' of order
$\left(n_S - 1\right)^2$; similarly, the running of the running can easily be
seen to be $\propto \left( n_S - 1 \right)^3$. This is true in nearly all
inflationary scenarios that have a scale--dependent spectral index.

Eq.(\ref{sol2}) can be solved for $g/c$. Bringing the denominator to the
left--hand side leads to a quadratic equation, which has two solutions. They
can be written as:
\begin{equation} \label{sol3}
\frac{g}{2c} = - \frac { \left( L+1 \right) \left( L^2 + 3L + 1 \right) + r_S L
  \left( 1.5 L + 2 \right) 
\pm L \sqrt{r_S \left[ \left( L+1 \right)^2 + 1 \right] + r_S^2 \left( 0.5L +
  1 \right)^2 } }
{\left( L^2 + 3L + 1 \right)^2 + r_S L \left( 2L + 3 \right) \left( L+1
  \right) } \,,
\end{equation}
where we have introduced the quantity
\begin{equation} \label{rs}
r_S \equiv \frac { \left( n_S - 1 \right)^2 } {2 \alpha_S} \,.
\end{equation}
Since $g$ and $c$ are real quantities, eq.(\ref{sol3}) only makes sense if the
argument of the square root is non--negative. Note that the coefficients
multiplying $r_S$ and $r_S^2$ inside the square root are both
non--negative. This means that the model can in principle accommodate any
non--negative value of $r_S$. However, small negative values of $r_S$ cannot
be realized. It is easy to see that the constraint on $r_S$ is weakest for
$L=0$. The argument of the square root is then positive if $2 r_S + r_S^2 >
0$, which implies either $r_S > 0$ or $r_S < -2$. Recalling the definition
(\ref{rs}) we are thus led to the conclusion
\begin{equation} \label{asmin}
\alpha_S \geq - \frac { \left(n_S - 1 \right)^2} {4}\,;
\end{equation}
this bound should hold on all scales, as long as the potential is described by
eq.(\ref{pot3}). Note that it is identical to the bound found in
ref.\cite{clmo}, i.e. it is not affected by adding the term $\propto L^2$
to the inflaton potential. This is somewhat disappointing, since recent data
indicate that $\alpha_S$ is negative at CMB scales. Even the generalized
version of the running mass model therefore cannot reproduce the current $1 \,
\sigma$ range of $\alpha_S$.

However, at the $2 \, \sigma$ level significantly positive $\alpha_S$ values
are still allowed. Let us therefore continue with our analysis, and search for
combinations of parameters within the current $2 \, \sigma$ range that might
lead to significant PBH formation. Using eqs.(\ref{sol1}) and (\ref{sol3}) we
can use $n_S(k_0)-1, \ \alpha_S(k_0)$ and $L_0 \equiv \ln ( \phi_0 / \phi_*)$
as input parameters in the last eq.(\ref{nab-slo}) to evaluate
$\beta_S(k_0)$. This can then be inserted into eq.(\ref{nofr}) to see how
large the density perturbations at potential PBH scales are.

This numerical analysis is most easily performed at the ``pivot scale'', where
the errors on $n_S$ and $\alpha_S$ are essentially uncorrelated; at $k-$values
above (below) this scale, $n_S$ and $\alpha_S$ are correlated (anticorrelated)
\cite{correlation}. The pivot scale for the ``WMAP7+BAO+$H_0$'' data set we
are using is $k_0 \equiv k_{\rm pivot} = 0.0155$ Mpc$^{-1}$
\cite{WMAP7}.\footnote{This is the smaller of two $k_{\rm pivot}$ values given
  in ref.\cite{WMAP7}. The difference between these two values is not
  important for our numerical analysis.} At this scale, observations give
\cite{WMAP7}: 
\begin{eqnarray} \label{range}
n_S(k_0) & = & 0.964 \pm 0.013 \, ; \nonumber \\
\alpha_S(k_0) & = & -0.022 \pm 0.020\, .
\end{eqnarray}

We saw above that requiring the correct normalization of the power spectrum at
CMB scales does not impose any constraint on the spectral parameters. However,
the model also has to satisfy several consistency conditions. To begin with,
it should provide a sufficient amount of inflation. In the slow--roll
approximation, the number of e--folds of inflation following from the
potential (\ref{pot3}) is given by:
\begin{eqnarray} \label{NofL}
\Delta N(L) &=& - \frac {2} {\tilde c} \int_{L_0}^L \frac {d \, L^\prime}
{L^\prime \left[ 1 + \frac {g}{2c} \left( L^\prime + 1 \right) \right] }
\nonumber \\
& = & - \frac {2} {\tilde c \left( 1 + \frac {g}{2c} \right) }
\left[ \ln \frac {L} {L_0} - \ln \frac {1 + \frac {g}{2c} (L+1)}
{ 1 + \frac {g}{2c} (L_0+1) } \right]\,,
\end{eqnarray}
where we have introduced the dimensionless quantity
\begin{equation} \label{ct}
\tilde c \equiv \frac {2 c M^2_{\rm P}} {V_0}\,;
\end{equation}
recall that it can be traded for $n_S(k_0)-1$ using eq.(\ref{sol1}). Moreover,
$L \equiv \ln (\phi / \phi_*)$ can be related to the scale $k$ through
\begin{equation} \label{kofL}
k(L) = k_0 {\rm e}^{\Delta N(L)}\,.
\end{equation}
This can be inverted to give
\begin{equation} \label{Lofk}
L(k) = L_0 \frac { E(k) \left( 1 + \frac {g}{2c} \right) }
{ 1 + \frac {g}{2c} \left[ L_0 \left( 1 - E(k) \right) + 1 \right] }\,,
\end{equation}
where we have introduced
\begin{equation} \label{E}
E(k) = \left( \frac {k}{k_0} \right)^{-\frac{\tilde c}{2} \left( 1 + \frac
  {g}{2c} \right)} \, .
\end{equation}
The problem is that the denominator in eq.(\ref{Lofk}) vanishes for some
finite value of $k$. This defines an extremal value of $\Delta N$:
\begin{equation} \label{DNex}
\Delta N_{\rm ex} = - \frac {2} { \tilde c \left( 1 + \frac{g}{2c} \right) }
\ln \left[ 1 + \frac {1}{L_0} \left( 1 + \frac {2c}{g} \right) \right]\,.
\end{equation}
A negative value of $\Delta N_{\rm ex}$ is generally not problematic, since
only a few e--folds of inflation have to have occurred before our pivot scale
$k_0$ crossed out of the horizon. However, a small positive value of $\Delta
N_{\rm ex}$ would imply insufficient amount of inflation after the scale $k_0$
crossed the horizon. In our numerical work we therefore exclude scenarios with
$-5 < \Delta N_{\rm ex} < 50$, i.e. we (rather conservatively) demand that at
least 50 e--folds of inflation can occur after $k_0$ crossed out of the
horizon.

A second consistency condition we impose is that $|L|$ should not become too
large. Specifically, we require $|L(k)| < 20$ for all scales between $k_0$ and
the PBH scale. For (much) larger values of $|L|$ our potential (\ref{pot3})
may no longer be appropriate, i.e. higher powers of $L$ may need to be
included.

Note that eq.(\ref{Lofk}) allows to compute the effective spectral index
$n_S(k)$ {\em exactly}:
\begin{equation} \label{nsofk}
n_S(k) - 1 = \left[ n_S(k_0) - 1 \right] \frac { L(k) + 1 + \frac {g}{2c}
  \left[ L(k)^2 + 3 L(k) + 1 \right] } {L_0 + 1 + \frac{g}{2c} \left( L_0^2 +
  3 L_0 + 1 \right) }\,.
\end{equation}
This in turn allows an {\em exact} (numerical) calculation of the spectral
index $n(k)$:
\begin{equation} \label{nofk}
n(k) - 1 = \frac {1} {\ln \frac {k}{k_0}} \int_0^{\ln \frac{k}{k_0}} \left[
  n_S(k^\prime) -1 \right] d \, \ln k^\prime \, .
\end{equation}

In our numerical scans of parameter space we noticed that frequently the exact
value for $n(k)$ at PBH scales differs significantly from the values predicted
by the expansion of eq.(\ref{nofr}); similar statements apply to
$n_S(k)$. This is not very surprising, given that $|\ln(k_0 R)| = 39.1$ for
our value of the pivot scale and $M_{\rm PBH} = 10^{15}$g. In fact, we noticed
that even if $\alpha_S(k_0)$ and $\beta_S(k_0)$ are both positive, $n_S(k)$
may not grow monotonically with increasing $k$. In some cases $n_S(k)$
computed according to eq.(\ref{nsofk}) even becomes quite large at values of
$k$ some 5 or 10 e--folds below the PBH scale. This is problematic, since our
calculation is based on the slow--roll approximation, which no longer works if
$n_S-1$ becomes too large. We therefore demanded $n_S(k) < 2$ for all scales
up to the PBH scale; the first eq.(\ref{nab}) shows that this corresponds to
$\eta < 0.5$. This last requirement turns out to be the most constraining one
when looking for combinations of parameters that give large $n(k_{\rm PBH})$.

\section{Numerical Results}

\begin{figure}[t]
\centering\includegraphics[angle=270,width=1.\textwidth]{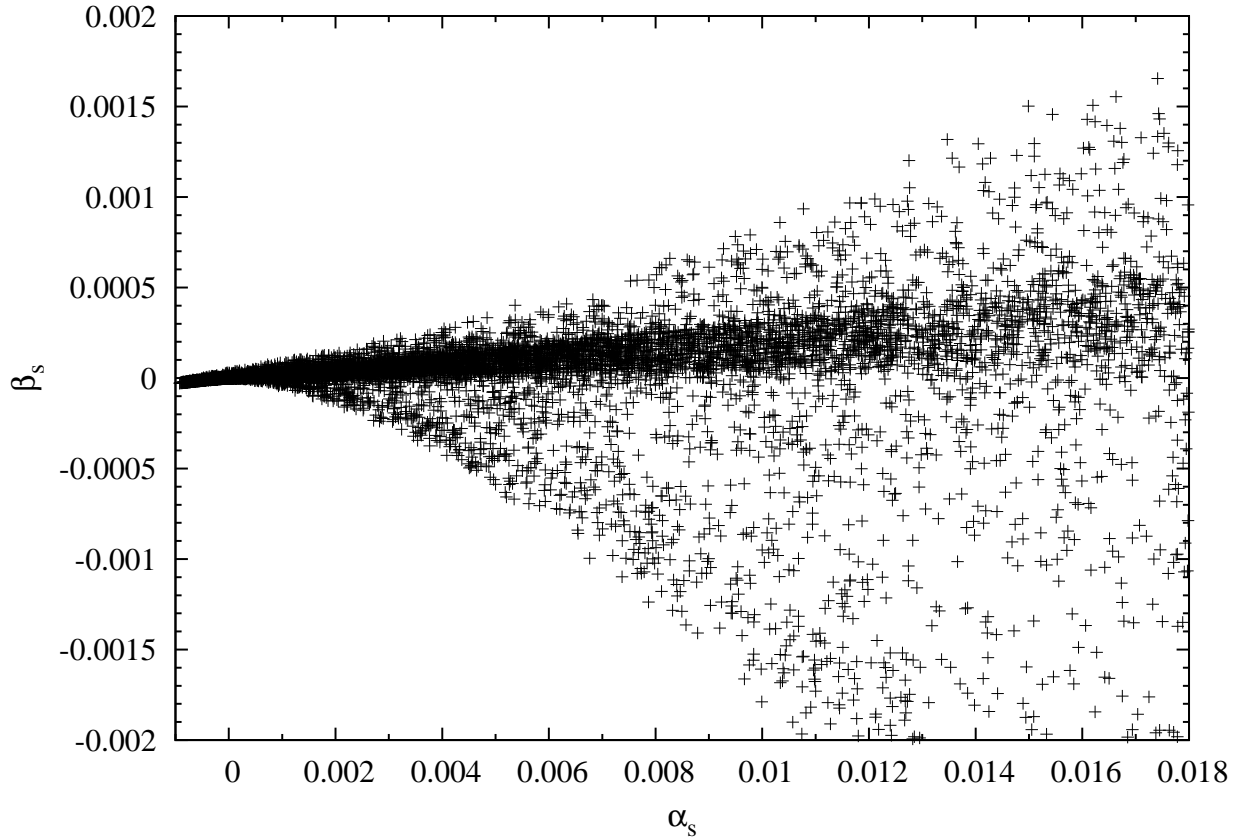}
\caption{Scatter plot of $\beta_S(k_0)$ vs. $\alpha_S(k_0)$. Here the model
  parameters $\tilde c \equiv 2 c M_P^2 / V_0, \ g/c$ and $L_0$ are scanned
  randomly, with flat probability distribution functions.}
\label{fig:Fig2}
\end{figure}

We are now ready to present some numerical results. We begin in
figure~\ref{fig:Fig2}, which shows a scatter plot of the spectral parameters
$\alpha_S(k_0)$ and $\beta_S(k_0)$, which has been obtained by randomly
choosing model parameters $\tilde c$ [defined in eq.(\ref{ct})], $g/c$ and
$L_0$ in the ranges\footnote{We actually only find acceptable solutions for
  $|\tilde c| < 0.8$.} $|\tilde c| \leq 1, \ |g| M_{\rm P}^2/V_0 \leq 1,
\ |L_0| \leq 20$. We require that $n_S(k_0)$ and $\alpha_S(k_0)$ lie within
their $2 \, \sigma$ ranges, and impose the consistency conditions discussed
above. The plot shows a very strong correlation between $\beta_S$ and
$\alpha_S$: if the latter is negative or small, the former is also small in
magnitude.  Moreover, there are few points at large $\alpha_S$, and even there
most allowed combinations of parameters lead to very small $\beta_S$. The
accumulation of points at small $\alpha_S$ can be understood from our earlier
result (\ref{sol2}), which showed that $\alpha_S$ is naturally of order
$(n_S-1)^2 < 0.004$ within $2\, \sigma$. Moreover, $\beta_S$ is naturally of
order $\alpha_S^{3/2}$. On the other hand, for $\alpha_S$ values close to the
upper end of the current $2 \, \sigma$ range, we do find some scenarios where
$\beta_S$ is sufficiently large to allow the formation of $10^{15}$ g PBHs.

We also explored the correlation between $n_S(k_0)$ and $\alpha_S(k_0)$ (not
shown). Here the only notable feature is the lower bound (\ref{asmin}) on
$\alpha_S(k_0)$; values of $\alpha_S(k_0)$ up to (and well beyond) its
observational upper bound can be realized in this model for any value of
$n_S(k_0)$ within the presently allowed range. Similarly, we do not find any
correlation between $n_S(k_0)$ and $\beta_S(k_0)$. This lack of correlation
can be explained through the denominator in eq.(\ref{sol2}), which also
appears (to the third power) in the expression for $\beta_S$ once
eq.(\ref{sol1}) has been used to trade $c$ for $n_S-1$: this denominator can
be made small through a cancellation, allowing sizable $\alpha_S$ even if
$n_S$ is very close to 1. Since the same denominator appears (albeit with
different power) in the expressions for $\alpha_S$ and $\beta_S$, it does not
destroy the correlation between these two quantities discussed in the previous
paragraph.
\begin{figure}[t]
\centering\includegraphics[angle=270,width=1.\textwidth]{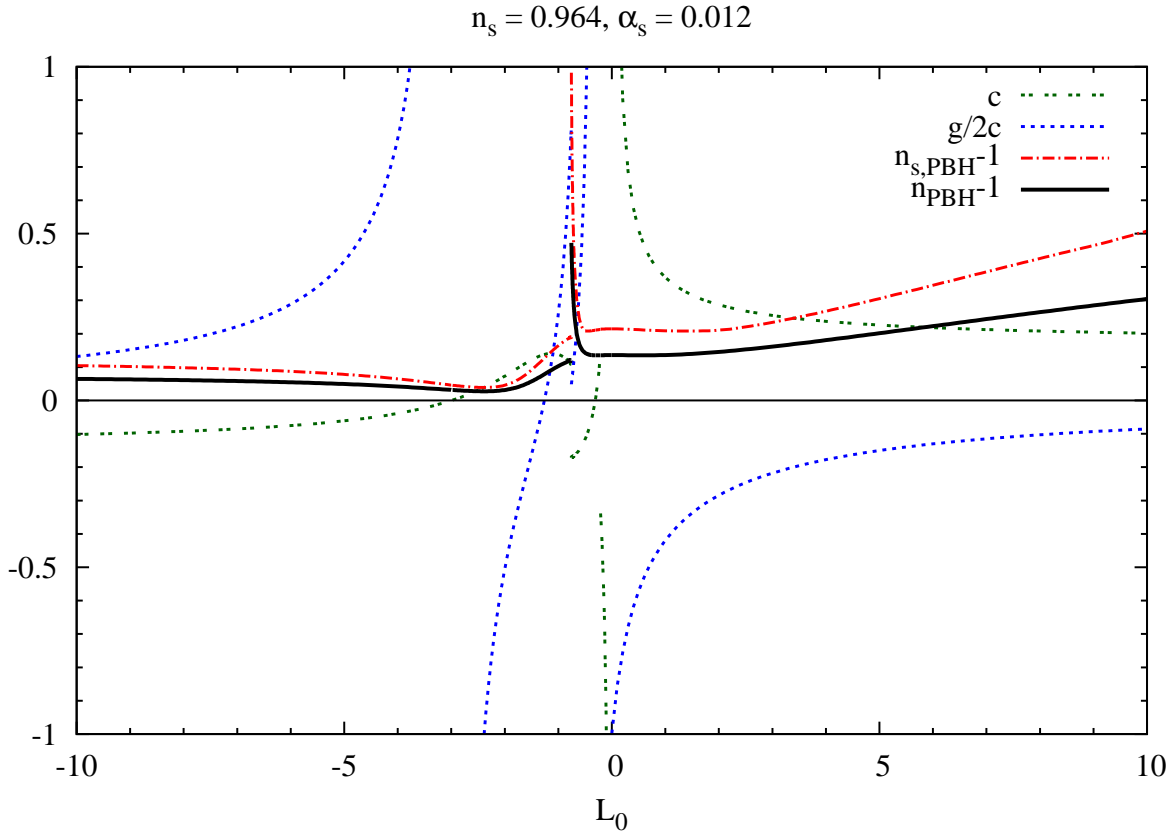}
\caption{The potential parameters $c$ (in units of $V_0/(2 M_{\rm P}^2)$;
 double--dotted [green] curve), $g/(2c)$ (dotted, blue), the effective
 spectral index $n_S-1$ at the PBH scale (dot--dashed, red) and the spectral
 index $n-1$ at the PBH scale (solid, black) are shown as functions of $L_0 =
 \ln ( \phi_0 / \phi_*)$, for $n_S(k_0) = 0.964$ and $\alpha_S(k_0) =
 0.012$. If both solutions in eq.(\ref{sol3}) for $g/c$ are acceptable, we
 have taken the one giving larger $n$ at the PBH scale.}
\label{fig:Fig3}
\end{figure}

Figure~\ref{fig:Fig2} indicates that large values of $n$ at the PBH scale can be
achieved only if $\alpha_S$ at the CMB scale is positive and not too small. In
figure~\ref{fig:Fig3} we therefore explore the dependence of the potential
parameters, and of the spectral parameters at the PBH scale on $L_0$, for
$n_S(k_0) = 0.964, \ \alpha_S(k_0) = 0.012$. Note that varying $L_0$ also
changes the parameters $c$ and $g$ (or $g/c$), see eqs.(\ref{sol1}) and
(\ref{sol3}). The latter in general has two solutions; however, for most
values of $L_0$, only one of them leads to sufficient inflation while keeping
$|L| < 20$; if both solutions are allowed, we take the one giving a larger
spectral index at the PBH scale, taken to be $1.5\times10^{15}$ Mpc$^{-1}$
corresponding to $M_{\rm PBH} = 10^{15}$ g. Note that $n_S$ and $n$ at the PBH
scale are calculated exactly, using eqs.(\ref{nab}) and (\ref{nofk}). We find
that the expansion (\ref{nofr}) is frequently very unreliable, e.g. giving the
wrong sign for $n-1$ at the PBH scale for $L_0 < -1$.

Figure~\ref{fig:Fig3} shows that $\tilde c$ is usually well below 1, as expected
from the fact that $n_S-1 \propto \tilde c$, see the first eq.(\ref{nab-slo}).
Moreover, in most of the parameter space eq.(\ref{sol3}) implies $|g| < |c|$;
recall that this is also expected, since $g$ is a two--loop term. We find $|g|
> |c|$ only if $|c|$ is small. In particular, the poles in $g/(2c)$ shown in
figure~\ref{fig:Fig3} occur only where $c$ vanishes; note that the spectral
parameters remain smooth across these ``poles''.

There are a couple of real discontinuities in figure~\ref{fig:Fig3}, where the
curves switch between the two solutions of eq.(\ref{sol3}). The first occurs
at $L_0 \simeq -0.756$. For smaller values of $L_0$, the solution giving the
smaller $|g/c|$ violates our slow--roll condition $|n_S-1| < 1$ at scales
close to the PBH scale. For larger $L_0$ this condition is satisfied. Just
above the discontinuity, where $n_S$ is close to 2 at the PBH scale, we find
the largest spectral index at the PBH scale, which is close to 0.47. Recall
from figure~\ref{fig:Fig1} that this will generate sufficiently large density
perturbations to allow the formation of PBHs with $M_{\rm PBH} = 10^{15}$
g. However, the formation of PBHs with this mass is possible only for a narrow
range of $L_0$, roughly $-0.756 \leq L_0 \leq -0.739$.

At $L_0 =-0.31$, $c$ goes through zero, giving a pole in $g/c$ as
discussed above. Then, at $L_0 = -0.214$, the second discontinuity
occurs. Here the curves switch between the two solutions of eq.(\ref{sol3})
simply because the second solution gives a larger spectral index at the PBH
scale. Right at the discontinuity both solutions give the same spectral index,
i.e. the curve depicting $n_{\rm PBH}$ remains continuous; however, $\tilde c$
jumps from about $0.132$ to $-0.339$. The effective spectral index $n_S$ at
the PBH scale also shows a small discontinuity. Recall from our discussion of
eq.(\ref{nsofn}) that $n_S$ will generally be larger than $n$ at the PBH
scale, but the difference between the two depends on the model parameters.

For very small values of $|L_0|$, $\tilde c$ becomes very large; this region
of parameter space is therefore somewhat pathological. For sizably positive
$L_0$, $n$ at the PBH scale increases slowly with increasing $L_0$, while
$|\tilde c|$ and $|g/c|$ both decrease. However, the spectral index at the PBH
scale remains below the critical value for the formation of long--lived PBHs.

Note that $L$ always maintains its sign during inflation, since $L=0$
corresponds to a stationary point of the potential, which the (classical)
inflaton trajectory cannot cross.  For most of the parameter space shown in
figure~\ref{fig:Fig3}, $|L|$ decreases during inflation. If $L_0 < 0$ decreasing
$|L|$ corresponds to $V'(\phi_0) < 0$, i.e. the inflaton rolls towards a
minimum of the potential at $\phi = \phi_*$. For $L_0 > 0$ we instead have
$V'(\phi_0) > 0$, i.e. the inflaton rolls away from a maximum of the
potential.

\begin{figure}[h!]
\centering{\includegraphics[angle=270,width=0.48\textwidth]{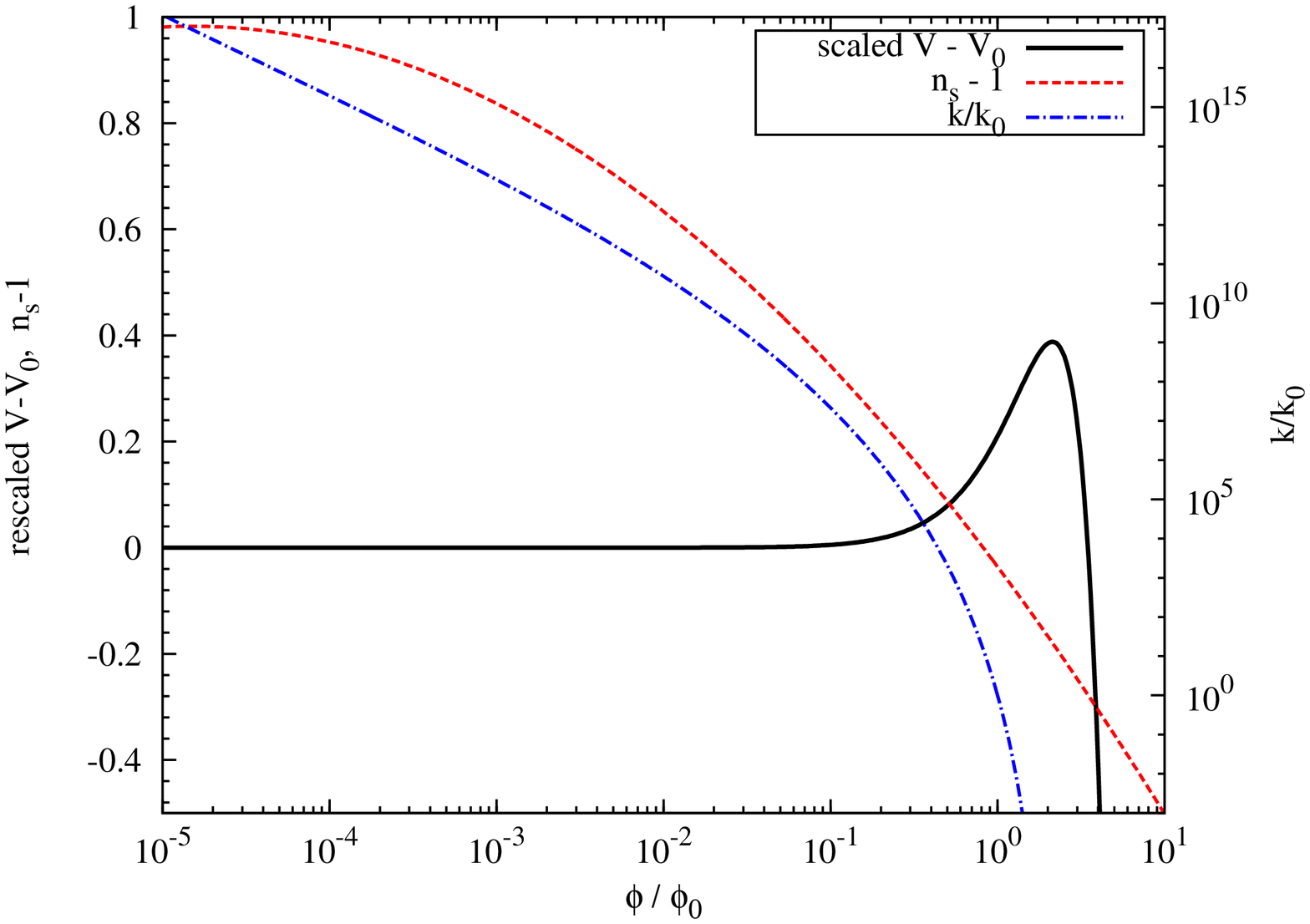}
\includegraphics[angle=270,width=0.48\textwidth]{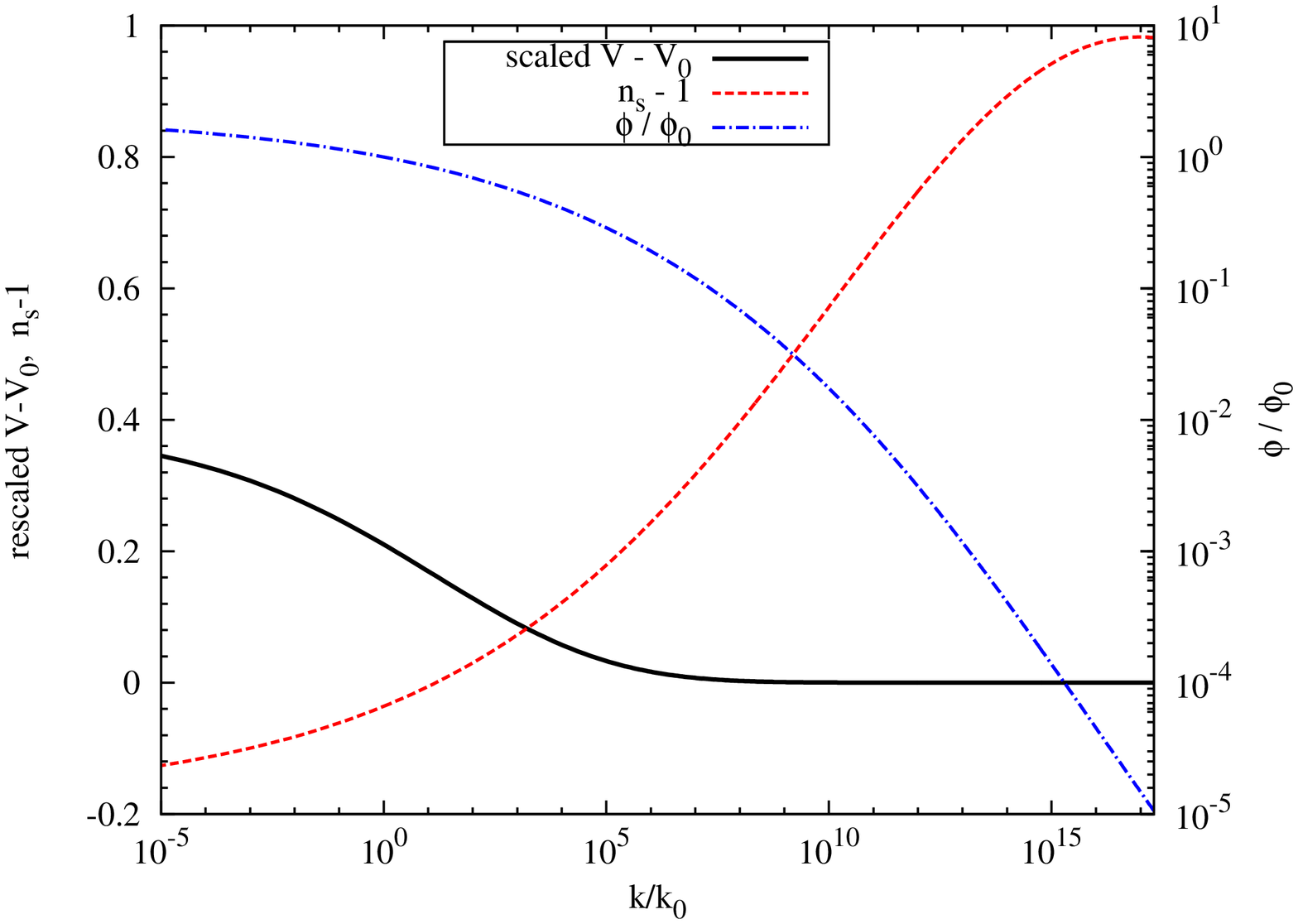}}
\caption{Evolution of the rescaled inflaton potential, $4 M_{\rm P}^2 (V-V_0)
  / (V_0 \phi_0^2)$ (solid, black), and of the effective spectral index $n_S$
  (dashed, red), as a function of the inflaton field $\phi/\phi_0$ (left
  frame) or of the ratio of scales $k/k_0$ (right frame). In the left frame
  the dot--dashed (blue) curve shows $k/k_0$, whereas in the right frame it
  depicts $\phi/\phi_0$; this curve in both cases refers to the scale to the
  right. We took $\tilde c = -0.1711, \ g/c = 0.09648, \ L_0 = -0.756$; these
  parameters maximize the spectral index at the PBH scale for $n_S(k_0) =
  0.964, \ \alpha_S(k_0) = 0.012$ (see figure~\ref{fig:Fig3}).}
\label{fig:Fig4}
\end{figure}

In fact, this latter situation also describes the branch of
figure~\ref{fig:Fig3} giving the largest spectral index at the PBH scale; since
here $L_0 < 0$, $|L|$ increases during inflation on this branch. This is
illustrated in figures~\ref{fig:Fig4}, which show the (rescaled) inflaton
potential as well as the effective spectral index as function of either the
inflaton field (left frame) or of the scale $k$ (right frame). Note that all
quantities shown here are dimensionless, and are determined uniquely by the
dimensionless parameters $\tilde c$ defined in eq.(\ref{ct}), $g/c$ and
$L_0$. This leaves two dimensionful quantities undetermined, e.g. $V_0$ and
$\phi_*$; one combination of these quantities can be fixed via the
normalization of the CMB power spectrum, leaving one parameter undetermined
(and irrelevant for our discussion).

The left frame shows that the (inverse) scale $k$ first increases quickly as
$\phi$ rolls down from its initial value $\phi_0$. This means that $\phi$
initially moves rather slowly, as can also be seen in the right frame. Since
$\alpha_S > 0$, the effective spectral index increases with increasing
$k$. The right frame shows that this evolution is quite nonlinear, although
for $k/k_0 \lsim 10^{10}$, $n_S(k)$ is to good approximation a parabolic
function of $\ln(k/k_0)$. However, for even smaller scales, i.e. larger $k$,
the rate of growth of $n_S$ decreases again, such that $n_S-1$ reaches a value
very close to $1$ at the scale $k = 1.6\times10^{16} \, k_0$ relevant for the
formation of PBHs with $M_{\rm PBH} = 10^{15}$ g. Recall that we only allow
solutions where $n_S(k) < 2$ for the entire range of $k$ considered;
figures~\ref{fig:Fig4} therefore illustrate our earlier statement that this
constraint limits the size of the spectral index at PBH scales.

The left frame of figure~\ref{fig:Fig4} shows that the inflaton potential as
written becomes unbounded from below for $\phi \rightarrow +\infty$. This can
be cured by introducing a quartic (or higher) term in the inflaton potential;
the coefficient of this term should be chosen sufficiently small not to affect
the discussion at the values of $\phi$ of interest to us. Note also that this
pathology of our inflaton potential is not visible in the right frame, since
assuming $\phi = \phi_0 < \phi_*$ at $k=k_0$, the inflaton field can never
have been larger than $\phi_*$: as noted above, it cannot have moved across
the maximum of the potential.

\begin{figure}[h!]
\centering\includegraphics[angle=270,width=1.\textwidth]{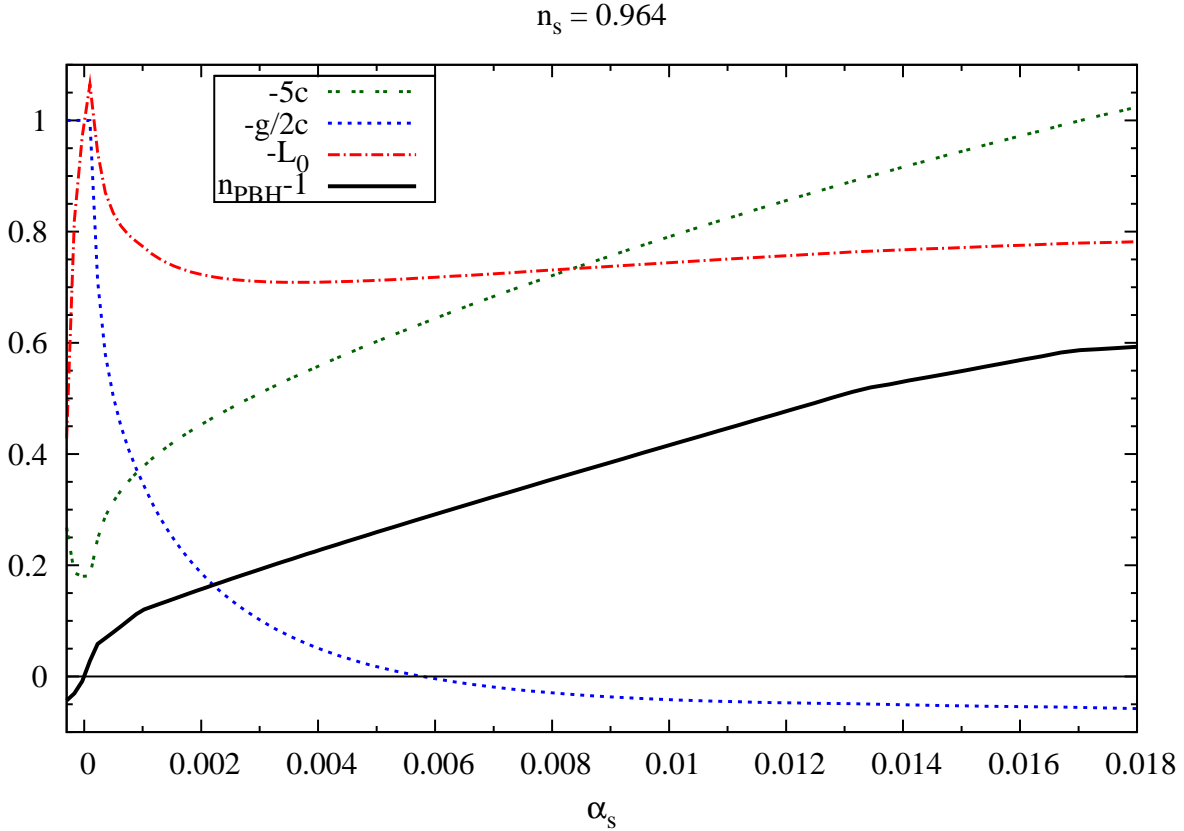}
\caption{The solid (black) curve shows the maximal spectral index at the PBH
  scale $k_{\rm PBH} = 1.6\times10^{19} k_0$ that is consistent with the
  constraints we impose, as function of $\alpha_S(k_0)$. The other curves show
the corresponding model parameters: $\tilde c$ (double-dotted, green),
multiplied with $-5$ for ease of presentation; $-g/(2c)$ (dotted, blue); and
$-L_0$ (dot--dashed, red). $n_S(k_0)$ is fixed at 0.964, but the bound on
$n_{\rm PBH}$ is almost independent of this choice.}
\label{fig:Fig5}
\end{figure}

Figure~\ref{fig:Fig2} indicated a strong dependence of the maximal spectral
index at PBH scales on $\alpha_S(k_0)$. This is confirmed by
figure~\ref{fig:Fig5}, which shows the maximal possible $n(k_{\rm PBH})$
consistent with our constraints as function of $\alpha_S(k_0)$, as well as the
corresponding values of the parameters $\tilde c, \ g/c$ and $L_0$. We saw in
the discussion of eq.(\ref{sol3}) that $\alpha_S < 0$ is only allowed for a
narrow range of $L_0$. In this very constrained corner of parameter space,
$n(k_{\rm PBH})$ remains less than $1$, although the effective spectral index
$n_S(k_{\rm PBH})$ can exceed $1$ for $\alpha_S(k_0) > -1.7\times10^{-4}$;
recall that for the given choice $n_S(k_0) = 0.964$, solutions only exist if
$\alpha_S(k_0) > -3.24\times10^{-4}$, see eq.(\ref{asmin}).

For $\alpha_S(k_0) \leq 1.5\times10^{-4}$ the optimal set of parameters lies
well inside the region of parameter space delineated by our
constraints. $n_S(k_{\rm PBH})$ therefore grows very fast with increasing
$\alpha_S(k_0)$. For these very small values of $\alpha_S(k_0)$, the largest
spectral index at PBH scales is always found for $g = -2c$, which implies that
the second derivative of the inflaton potential also vanishes at $\phi =
\phi_*$, i.e. $\phi_*$ corresponds to a saddle point, rather than an extremum,
of the potential.

For slightly larger values of $\alpha_S(k_0)$ the choice $g = -2c$ allows less
than 50 e--folds of inflation after $k_0$ exited the horizon. Requiring at
least 50 e--folds of inflation therefore leads to a kink in the curve for
$n(k_{\rm PBH})$. The optimal allowed parameter set now has considerable
smaller $|g/c|$, but larger $|\tilde c|$.

The curve for the maximal $n(k_{\rm PBH})$ shows a second kink at
$\alpha_S(k_0) = 0.001$. To the right of this point the most important
constraint is our requirement that $|n_S-1| < 1$ at all scales up to $k_{\rm
  PBH}$, as discussed in connection with figures~\ref{fig:Fig3} and
\ref{fig:Fig4}. Note that for $\alpha_S(k_0) > 0.014$, the optimal parameter
choice leads to $n_S$ reaching its maximum at some intermediate $k$ close to,
but smaller than, $k_{\rm PBH}$. This leads to a further flattening of the
increase of $n(k_{\rm PBH})$.

We nevertheless see that for values of $\alpha_S(k_0)$ close to the upper end
of the $2\, \sigma$ range specified in (\ref{range}) the spectral index at the
scale relevant for the formation of $10^{15}$ g PBHs can be well above the
minimum for PBH formation found in section~2. Figure~\ref{fig:Fig1} then implies
that the formation of considerably heavier PBHs might be possible in running
mass inflation. However, larger PBH masses correspond to smaller $k_{\rm
  PBH}$, see eq.(\ref{R}). This in turn allows for less running of the
spectral index. In order to check whether even heavier PBHs might be formed
during the slow--roll phase of running mass inflation, one therefore has to
re--optimize the parameters for different choices of $k_{\rm PBH}$.

\begin{figure}[t]
\centering\includegraphics[angle=270,width=1.\textwidth]{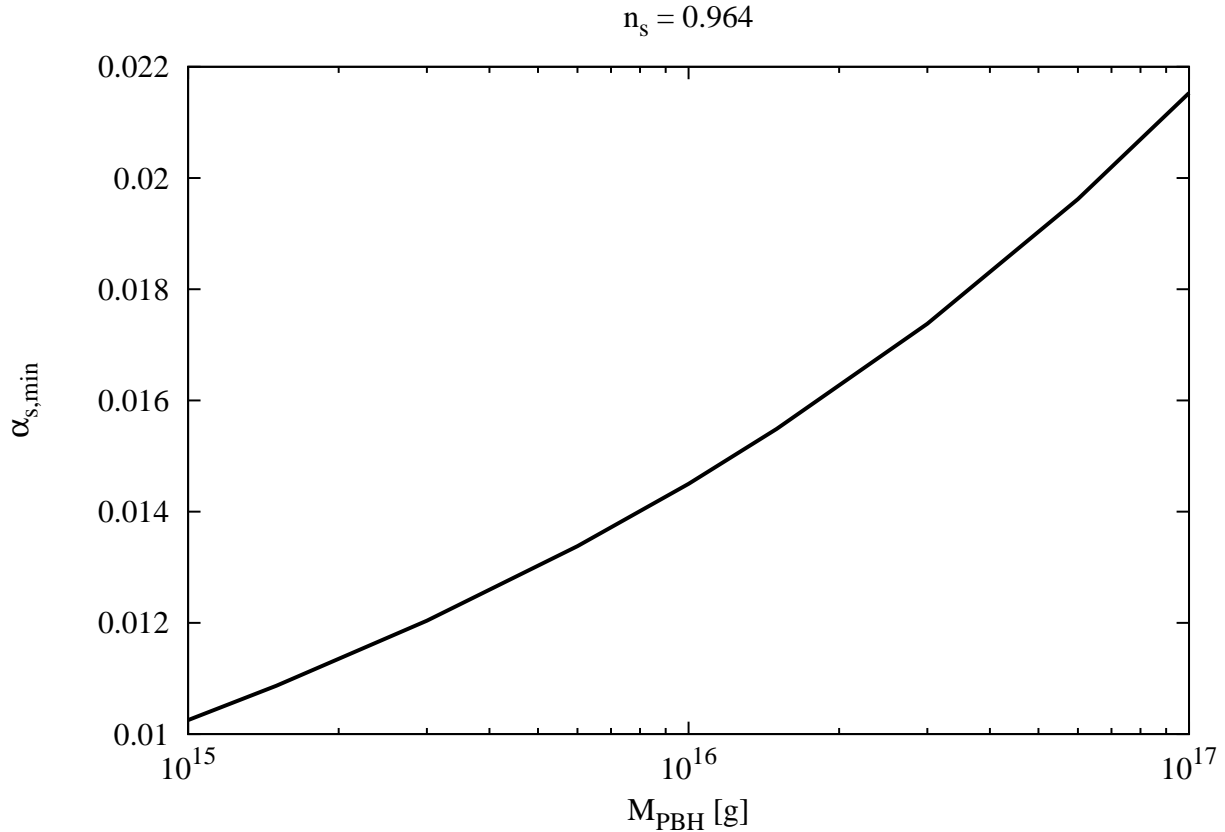}
\caption{The minimal value of $\alpha_S(k_0)$ that allows the formation of
 primordial black holes of a given mass as a result of density perturbations
 produced during the slow--roll phase of inflation.}
\label{fig:Fig6}
\end{figure}
The result of such an analysis is shown in figure~\ref{fig:Fig6}. We see that in
the given model, the formation of PBHs that are sufficiently massive, and
hence long--lived, to be CDM candidates could only have been triggered during
the slow--roll phase of inflation if $\alpha_S(k_0)$ is more than one standard
deviation above its central value. If $\alpha_S(k_0)$ is at the upper end of
its present $2\, \sigma$ range, PBHs with mass up to $5\times10^{16}$ g could
have formed as result of density perturbations created at the end of
slow--roll inflation. These results are again insensitive to the value of
$n_S(k_0)$.

\section{Summary and Conclusions}

\paragraph{}

In this paper we have investigated the formation of primordial black holes in
the radiation--dominated era just after inflation. We have focused on density
perturbations originating from the slow--roll phase of inflation. In section~2
we reviewed the Press--Schechter type formalism for PBH formation. We found
that the formation of PBHs with mass larger than $10^{15}$ g, which could form
(part of) the cold Dark Matter in the Universe, the spectral index at scale
$k_{\rm PBH}$ should be at least $1.37$, even for the lower value of $1/3$
for the threshold $\delta_{\text{th}}$. This value is higher that the value
$1.25$ found in \cite{Green2} because here we have assumed that the mass of
the collapsed region to form PBH is only 20\% of the entire energy density
inside the particle horizon.

This spectral index is much above the value measured in the CMB. PBH formation
therefore requires significant positive running of the spectral index when $k$
is increased. In contrast, current observations favor a negative first
derivative of the spectral index. For the central value $n_S(k_0) = 0.964$,
where $k_0$ is the CMB pivot scale, the first derivative $\alpha_S(k_0)$ would
need to exceed $0.020$ if it alone were responsible for the required increase
of the spectral index; this is outside the current $2\,\sigma$ range for this
quantity. However, the second derivative (the ``running of the running'') is
currently only very weakly constrained. We showed in a model--independent
analysis that this easily allows values of $n(k_{\rm PBH})$ large enough for
PBH formation, even if the first derivative of the spectral index is negative
at CMB scales.

In section~3 we applied this formalism to the ``running--mass'' model of
inflation, which had previously been shown to allow sufficiently large
positive running of the spectral index to yield PBH formation.  In contrast to
previous analyses, we included a term quadratic in the logarithm of the field,
i.e. we included the ``running of the running'' of the inflaton mass along
with the ``running of the running'' of the spectral index.  We showed that
this model can accommodate a sizably positive second derivative of the
spectral index at PBH scales. However, this is only possible if the first
derivative is also positive and sizable. In fact, like most inflationary
scenarios with a smooth potential \cite{Peiris} the model does not permit
large negative running of the spectral index at CMB scales. Current data prefere negative running of the spectral index at CMB scales, but are consistent with sufficiently large positive running at the $2 \, \sigma$ level. Moreover, we saw that a quadratic (in $\ln k$) extrapolation of the spectral index to PBH
scales is not reliable, and therefore computed the spectral index
exactly. Imposing several consistency conditions, we found that density
perturbations that are sufficiently large to trigger PBH formation only occur
for a very narrow region of parameter space. Among other things, the signs of
the parameters of the inflaton potential must be chosen such that the
potential has a local maximum, and the initial value of the field must be
slightly (by typically less than one e--fold) below this maximum.

We emphasize that obtaining sufficiently large density perturbations at small
scales is only a necessary condition for successful PBH formation. One major
challenge of this model is that for parameters allowing PBH formation the
spectral index keep increasing at yet smaller scales. Parameters that lead to
the formation of many PBHs with mass around $10^{15}$ g, which could form the
Dark Matter in the universe, would predict the over--production of unstable
PBHs, in conflict with data e.g. from the non--observation of black hole
evaporation and (for yet smaller masses) Big Bang Nucleosynthesis. This
problem seems quite generic for this mechanism. One way to solve it might be
to abruptly cut off inflation just after the scales relevant for the formation
of the desired PBHs leave the horizon, which could e.g. be achieved by
triggering the waterfall field in hybrid inflation.  However, this is somewhat
in conflict with the use of the slow--roll formalism of structure formation,
which is usually assumed to require a few more e--folds of inflation after the
scales of interest left the horizon.Moreover, a sharpe end of inflation means that the visible universe only inflated by about $45$ e--folds; this solves the horizlatness problems only if the reheat temperature was rather low \cite{Turner Book}.

We can therefore not state with confidence that formation of PBHs as Dark
Matter candidates is possible in running--mass inflation; all we can say is
that certain necessary conditions can be satisfied. Even that is possible only
for a very limited range of parameters. This also means that constraints from
the over--production of PBHs only rule out a small fraction of the otherwise
allowed parameter space of this model.

\section*{Acknowledgments}

This work is supported in part by the DFG Transregio TR33 ``The Dark
Universe''. EE thanks the Bonn--Cologne Graduate School for support, and the
KIAS School of Physics for hospitality. MD thanks the particle theory group of
the university of Hawaii at Manoa for hospitality while this work was
completed.

\end{document}